\documentclass[12pt,preprint]{aastex}

\usepackage[usenames]{color}

\usepackage{graphicx}
\usepackage{epstopdf}
\usepackage{url}

\newcommand{\be}{\begin{equation}}
\newcommand{\ee}{\end{equation}}
\newcommand{\ba}{\begin{eqnarray}}
\newcommand{\ea}{\end{eqnarray}}
\newcommand{\bc}{\begin{center}}
\newcommand{\ec}{\end{center}}
\newcommand{\lsi}{LS~I~+61$^{\circ}$\,303}

\begin{document}

\shorttitle{\textsc{Lightcurve variability in the X-ray emission of \lsi}}
\shortauthors{\textsc{Torres et al.}}

\title{\textsc{Variability in the orbital profiles of the X-ray emission of the $\gamma
$-ray binary \lsi}}
\author{Diego F. Torres\altaffilmark{1,2},
Shu Zhang\altaffilmark{3}, Jian Li\altaffilmark{3}, Nanda Rea\altaffilmark{1}, 
G. Andrea Caliandro\altaffilmark{1},
Daniela Hadasch\altaffilmark{1},
Yupeng Chen\altaffilmark{3},
Jianmin Wang\altaffilmark{3,4} \&
Paul S. Ray\altaffilmark{5}}

\altaffiltext{1}{Institut de Ci\`encies de l'Espai (IEEC-CSIC),
              Campus UAB,  Torre C5, 2a planta,
              08193 Barcelona, Spain}
\altaffiltext{2}{Instituci\'o Catalana de Recerca i Estudis Avan\c{c}ats (ICREA).
              Email: dtorres@ieec.uab.es}
\altaffiltext{3}{Laboratory for Particle Astrophysics, Institute of High
Energy Physics, Beijing 100049, China}
\altaffiltext{4}{Theoretical Physics Center for Science Facilities (TPCSF), CAS, China}
\altaffiltext{5}{Space Science Division, Naval Research Laboratory, Washington, DC 
20375-5352}

\begin{abstract}
We report on the analysis of  \textit{Rossi X-ray Timing Explorer} (\textit{RXTE}) 
Proportional Counter Array (PCA) monitoring observations of the $\gamma$-ray binary system \lsi, 
covering  35 full cycles of its orbital motion. This constitutes the largest continuous 
X-ray monitoring dataset analyzed to date for this source. Such an extended analysis 
allows us to report: a) the discovery of variability in the orbital profiles of the X-
ray emission, b) the existence of a few (recent) short flares on top of the overall 
behavior typical of the source, which, given the PCA field-of-view, may or may not be 
associated with \lsi, and c) the determination of the orbital periodicity using soft X-ray data alone.
\end{abstract}


\section{Introduction}

{ 
  \lsi\ is one of the few high-mass X-ray binaries that have been recently detected at TeV (Albert et al. 2006) and GeV energies (Abdo et al. 2009).
Indeed, \lsi\ shows periodic TeV emission modulated by the orbital motion (Albert et al. 2009), with a TeV peak out of phase with that observed at GeV energies,  variable X-ray
emission  and radio counterpart (see below).
In the last few years, there has been a burst of activity trying to understand the nature of this source, particularly whether it is a pulsar or a black hole system, and which are the mechanisms that lead to the multi-wavelength behavior (e.g., Bednarek 2006, Dubus 2006, Gupta \& Bottcher 2006,  Sierpowska-Bartosik \& Torres 2009). In part, these efforts were prompted by the results of a  
large VLBI campaign (Dhawan et al.~2006), which discovered rapid
changes along the orbit 
in the orientation of what seems to be a
cometary tail, consistent with being the result of a pulsar wind. In fact, if due to a jet, the changing
morphology of the radio emission along the orbit would require a
highly unstable jet, and the shape of this structure would not be expected
to be reproduced orbit after orbit  (Albert
et al.~2008; see however Massi \& Bernad\'o 2009 for an alternative explanation).
The repeatability of the orbital modulation and \textit{simultaneous} multiwavelength studies are thus crucial for constraining models.
}

However, soft X-ray pointed observations of \lsi\ have in general been too limited to cover full 
orbits of the system, which has a period of $\sim$26.5 days, or to study long-term 
evolution of the X-ray orbital profile. 
As an example, the observations at soft X-rays conducted by  {\em XMM-Newton} (Neronov 
\& Chernyakova 2006; Chernyakova et al. 2006, Sidoli et al. 2006), {\em Chandra} 
(Paredes et al. 2007, Rea et al. 2010), {\em ASCA} (Leahy et al. 1997), {\em ROSAT} 
(Goldoni \& Mereghetti 1995; Taylor et al. 1996), and {\em Einstein} (Bignami et al. 
1981) were all too short to cover even a single full orbit.  
Longer term-observations of \lsi\ were performed using instruments such as the
\textit{Rossi X-ray Timing Explorer} (\textit{RXTE}) All Sky Monitor (ASM) (see Leahy 
2001)  and {\em Swift}-XRT (Esposito et al. 2007) and, at harder X-rays, by {\em 
INTEGRAL}-IBIS/ISGRI (Chernyakova et al. 2006, Zhang et al. 2010).
The most aggressive monitoring campaign at soft X-rays was done so far with the 
Proportional Counter Array (PCA) on {\it RXTE}.
Intensive monitoring during the month of March 1996 were analyzed by Harrison et al. 
(2000) and Grenier and Rau (2001).  Smith et al.\ (2009) recently analyzed about five 
months of {\it RXTE}/PCA observations covering the period 2007 August 28 to 
2008 February 2, reporting the discovery of several flares but only a marginal detection 
of orbital modulation. Furthermore, their orbital profile hinted at a two-peak 
lightcurve in the 2--10 keV band (see their Figure 2). Hint of a similar two-peak feature 
was reported by Paredes et al.\ (1997) with 10 months of {\it RXTE}/ASM data, covering 
1996 February to December. { Although the two, as yet unconfirmed, 
X-ray peaks are not located at the expected phases according to 
earlier proposed two-peaked accretion models (e.g. Taylor et al. 1992, Mart\'i \& Paredes 1996), such models do predict two local maxima in the radio and X-ray lightcurves.}

An alternative approach to the study of \lsi\ has been to focus on possible correlated 
variability, particularly after the system has been detected at higher energies (Albert 
et al.\ 2006, Abdo et al.\ 2009). The first {\it strictly  simultaneous} observations at 
TeV and X-rays of \lsi\ have been recently presented by the MAGIC collaboration (Albert 
et al. 2009). Using observations by  {\it XMM-Newton}, {\em Swift}, and MAGIC during $
\sim60\%$ of an orbit, it was found that there is a correlation between the X-ray and 
TeV emission at the time where the latter was measurable. Other campaigns, e.g., Acciari 
et al. (2009), fall short of achieving such a result, probably because of 
non-simultaneity or being based on short snapshots of the system. But, given that 
the soft X-ray and the TeV emission themselves can be highly variable individually, 
a simultaneous correlation may only be indicative of local physical conditions, and not 
taken as a proof of a persistent correlation between energy bands, orbit after orbit.  Indeed, at several recent conferences where results from the VERITAS array have been 
presented (e.g. Aliu et al. 2010), it was noted that \lsi\ could now be in an 
unusual TeV low state.

In this Letter we report on several years of sensitive X-ray monitoring of \lsi\ obtained by the {\it RXTE} satellite (see Gruber et al. 1996). The observations 
covered 35 full cycles of the 26.496 day
binary period and constitute the largest continuous pointed X-ray monitoring dataset on
\lsi\ to date.

\section{Observations and data analysis} 

Our dataset covers the period between 2006 October and 2010 March, and includes 326
{\it RXTE}/PCA pointed observations 
identified by proposal numbers 92418, 93017, 93100, 93101, 93102, 94102, and 95102. These 
observations provide a total of 483 ks of exposure time on the source.  Here, we  
focus on the data since 2007 September, which constitute the majority of the data (473 ks) and have smaller time gaps between observations. 

The analysis of PCA data was performed using HEASoft 6.6.   We filtered the data using 
the standard {\it RXTE}/PCA criteria.
To be precise, only PCU2 (in the 0--4 numbering scheme) has been used for the 
analysis, because it is the only PCU that was 100\% on during the observations. We select time intervals where the source elevation is $>10^{\circ}$ and the pointing offset is $<0.02^{\circ}$. 
The background file used in analysis of PCA data is 
the most recent available  from the HEASARC website for faint sources,
and detector breakdown events have been removed.\footnote{The background file is 
\url{pca_bkgd_cmfaintl7_eMv20051128.mdl} and see the website:
\url{http://heasarc.gsfc.nasa.gov/docs/xte/recipes/pca_breakdown.html}
for more information on the breakdowns. The data have been barycentered using the {\tt 
FTOOLS} routine {\tt faxbary} using the JPL DE405 solar system ephemeris. 
}


To search for a periodic signal in the lightcurve data, we used the Lomb-Scargle periodogram method.
Power spectra were generated for the light curve using the PERIOD subroutine  (Press \& 
Rybicki 1989), and checked using the {\tt powspec} tool of the {\tt XRONOS} software 
package.
The oversampling and high-frequency factors which set the period resolution and range 
(Press \& Rybicki 1989) were set 
to search for 
periodic variability in the 1 -- 1000 days range. The Lomb-Scargle technique allows us to 
sample below the average Nyquist period (with reduced sensitivity) due to the unevenly 
sampled nature of the data, without creating problems due to aliasing.
The 99\% white noise significance levels were estimated using Monte Carlo simulations 
(see e.g. Kong, Charles \& Kuulkers 1998). The 99\% red noise significance levels were 
estimated using the REDFIT subroutine, which can provide the red-noise spectrum via 
fitting a first-order auto-regressive process to the time series  (Schulz \& Mudelsee 
2002).\footnote{See ftp://ftp.ncdc.noaa.gov/pub/data/paleo/softlib/redfit}


\section{Results}

The upper panel of Figure 1 represents the lightcurve (in 64 seconds time bins in the 
3--30 keV band) of the whole extent of the observations analyzed.  This data set significantly extends (more than by a factor of 5) that considered by Smith et 
al.\ (2009), which covers only the 2007--2008 campaign, from MJD 54340 to 54498.   
From this lightcurve one can see 
that, on top of the overall behavior typical of the source, there 
are several short flares. A few of these short flares were also observed by Smith et al. 
(based on the same data). We observe additional flares at 
about MJD 54670.84 (first reported by Ray et al. 2008) and at MJD 54699.65. 
A detailed study of these 
flares will be provided elsewhere. These short flares would significantly affect the 
study of the orbital profile of \lsi , hence we decided to remove them from the current 
analysis. To cut them out,  clipped all observations with count rates above 3 times the mean value of 1.551 $\pm$ 0.008 counts s$^{-1}$ (see horizontal line in top panel of Figure 1). We tested that our results were not sensitive to changing the cut to 
5$\sigma$ above the mean, or cutting based on fitting a Gaussian standard deviation $\sigma$. With this flux cut,
we folded the data modulo the radio period of 26.4960$\pm$0.0028 days (Gregory 2002)
with the orbital zero phase taken at JD 2443366.775 (Gregory \& Taylor 1978). The result can be seen in the second panel of 
Figure 1. A clear orbital modulation can be seen in the X-ray data, with a peak rising 
up around phase 0.45. This phase is shortly after the periastron passage, which is between phase 0.23--0.30 (Casares et al. 2005; Aragona et al. 2009; Grundstrom et al. 2007).

The third panel of Figure 1 shows the power spectrum analysis of the unfolded 
lightcurve. A peak is found at 26.68$\pm$0.58 days,
matching the 
orbital period of the system and presenting a power density  of $\sim$160. 
This is the first time that such an X-ray orbital periodicity is clearly 
shown in PCA data. The white noise and the red noise, both at 99\% confidence level, are 
marked in the same Figure, and present a much smaller power. 
A second peak is visible in the power spectrum, corresponding 
to the first harmonic, with periodicity at 13.36$\pm$0.12 days.

We investigated whether the orbital light curve is stable in time. To that aim, 
we divided our data into 5 periods of 6 months each, which provided good statistics 
at all the phases of the orbit. Figure \ref{6months} shows the folded lightcurve in each of the 6-months periods, 
ordered by time. 
Figure \ref{2a} (left panel) shows a superposition of the folded lightcurves obtained in 
each 6-months period. Significant 
(more than 7$\sigma$) variations in the flux at a fixed phase value appear, with the 
largest differences being located in the phase range around the global peak 
seen in the middle panel of Figure 1.
In Figure \ref{2a} (right panel)  we plot the modulated flux fraction (obtained as $
(c_\mathrm{max}-c_\mathrm{min})/(c_\mathrm{max}+c_\mathrm{min})$, where $c_\mathrm{max}$ and $c_\mathrm{min}$ are the maximum and minimum count rates 
found in each 6-months orbital profile after background subtraction; see Figure \ref{6months}) as a function of time (one point for each of the 6-months period 
considered). It is immediately obvious that, while the orbital average value of the count rate remains similar, the modulated signal significantly varies in 
time, increasing monotonically during the observations.

{ From the individual 6-month orbital profiles (shown in Figure \ref{6months}), and the example of month-by-month 
variability (shown in Figure \ref{22}), one can conclude that the X-ray profile is not stable at any of these 
timescales. 
We find the phase of X-ray maximum anywhere from phase 0.3 to 0.9 in both the single orbit and 6-month averages.
Looking at the lightcurves, one sees changes both in flux level and shape, 
starting with the appearance of  a weakly-modulated double-peaked structure and evolving into 
a more clearly visible, broad single-peaked lightcurve at later times. 
However, fitting one and two sine functions having the orbital and 
half the orbital periods, respectively, we find large $\chi^2$-values compared to the number of d.o.f.
Then, in both cases, the fits are not good enough to make claims, indicating that more harmonics are needed, which can not be well tested with the 
current data given the low d.o.f.
In addition, the power spectrum remains featureless in most of these shorter 6-months periods due to limited statistics (i.e.,
periodicity hints remain below the red noise in most of the individual 6-months datasets). In one of the 6-month periods though (the third one), the X-ray periodicity is significantly seen in the power spectrum.
We also noted 
signals appearing between 1 and $\sim$ 2 days, which result from the spacing of the 
observations and are thus unrelated to the physical behavior of the source. }

{ We also report on several flares (see also Smith et al. 2009) }
observed in this long-term monitoring. However, caution should be exercised  ---the PCA 
field of view is about 1 degree (FWHM)
so there is no direct evidence for the relation of this flaring phenomenology with \lsi. 
Since  none of these short flares were observed by any instrument with a better 
spatial resolution,  we can not exclude that they were generated 
in a nearby (in sky projection) source.

Similarly, on 2008 September 10th, \textit{Swift}-BAT triggered on a short SGR-like burst from 
the direction of \lsi\ (de Pasquale et al. 2008). The BAT location of the burst had an 
uncertainty of 2.2 arcminutes (all at 90\% confidence level; Barthelmy et al. 2008), 
centered 88 arcseconds away from the sub-arcsecond accurate position of \lsi.
The other telescope onboard, \textit{Swift}-XRT did not detect the burst because it started observing 921s after the BAT trigger
(Evans et al. 2008). The XRT did detect \lsi\ (being the brightest source in the 
field) at a flux level and a spectrum consistent with what was previously observed by 
other missions and \textit{Swift}-XRT itself (Leahy et al. 1997, Greiner \& Rau 2001, Sidoli et 
al. 2006, Esposito et al. 2007). 
Following these notices Dubus \& Giebels (2008) noted that the 
burst location, lightcurve, duration, fluence, and spectra were fully consistent with a 
soft  $\gamma$-ray  repeater/anomalous X-ray pulsar short duration burst originating 
from \lsi. They then considered this to be the first manifestation of magnetar-like 
activity in a high-mass X-ray binary, concluding that \lsi\ was likely to host a young, 
but highly magnetized pulsar. This is a tantalizing hypothesis for which, like the 
association of the PCA flares with \lsi, there is yet no proof.
Rea \& Torres (2008) and Mu\~noz-Arjonilla et al. (2008) analyzed archival {\it Chandra} 
ACIS-I observation of \lsi\ performed on 2006 April 7th (50ks;  Paredes et al. 2007), 
showing that many faint sources are detected in the field of view. Thus, any of the 
several X-ray sources without a radio counterpart within the field of view could be the 
origin of the magnetar-like burst. The {\it RXTE} PCA observation nearest to the burst is 
ObsID 93102-01-32-01, MJD 54719 at 18:17 UT, about 6 hours after the burst. The total exposure 
time is 1492 s. No flares are detected in this observation. The max count rate is 1.86 
cts/s and the average count rate is 1.09 cts/s. This latter result is consistent with 
the report by Ray et al. (2008), made immediately after the Swift burst was announced.

\section{Discussion}

 The discovery of high and very high energy gamma-ray emitting X-ray binaries 
triggered an intense effort to understand the particle acceleration, absorption, and emission mechanisms 
in these binary systems. The eccentricity and relatively small orbital periods provide significant changes of physical conditions 
along the orbit. Despite this effort, the nature of the compact object and emission mechanisms powering these systems is not settled. 
Because of this, multi-frequency observations and long-term monitoring campaigns such as the one reported here are essential. Multi-frequency observations can provide knowledge of the dominance of single or multiple particle populations, and of the nature of these particles. For instance, if the radiation mechanisms are dominated by a single particle population, an X-ray/VHE correlation with similar fluxes would favor leptonic models over hadronic ones, where the luminosity in X-rays should be less, given their origin in secondary particles. On the other hand, long-term monitoring provides information on possible trends in the overall behavior of the source, and gives perspective as of the steadiness or variability of the former conclusions. Such monitoring can also catch unusual events or source states that could provide the key to the nature of the source. There are a few potential observations that would conclusively demonstrate the true nature, such as the discovery of pulsations at any frequency or the detection of clear accretion signals, both of  which are for the moment still elusive (see e.g., Rea et al. 2010).

This Letter shows that the soft X-ray emission from \lsi\ presents a periodic behavior 
(visible only for long integration times) at the orbital period, whose shape varies at all timescales explored. Profile variability 
is seen from orbit-to-orbit all the way up to multi-year timescales. The phase of the profile maximum also changes.
 It is then possible that the X-ray emission is orbitally modulated due to the interaction of a stellar wind flow with a pulsar wind (e.g., Dubus et al. 2006, Sierpowska-Bartosik \& 
Torres 2008) or a jet (e.g., Gupta  \& B\"ottcher 2006, Bosch-Ramon et al. 2006), but the details of such variability may strongly depend on the intrinsic behavior of the Be 
stellar wind present in the system. 
Our results imply the following: 
\begin{enumerate}

\item The study of short-term, simultaneous multifrequency observations such as 
the one made by the MAGIC collaboration (Albert et al. 2009) produce local-in-time 
information useful for determining the process or the primaries responsible for  the radiation detected, but do not establish a steady overall behavior.  { The fact that the X-ray and the TeV 
emission from \lsi\ are correlated in a single short observation cannot 
be used to claim that the 
position in phase of the maxima in the lightcurves are maintained on timescales longer 
than the simultaneous observations themselves. In fact, given the variable nature of the X-ray emission
a correlation found in a short observation might not be sufficient proof that this correlation maintains in time.}

\item Long-term variability in X-rays do occur on timescales of years. Thus, if the 
TeV and X-ray emission are indeed always correlated, the TeV maximum should also 
vary in phase. In addition, the relative strengths of the emission in these bands can be affected 
by the level of absorption at which the maximal photon production happens (e.g., see 
Sierpowska-Bartosik \& Torres 2009), varying from orbit to orbit.

\end{enumerate}

\acknowledgements
This work has been supported by grants AYA2009-07391
and SGR2009-811 and was further subsidized by the National Natural Science 
Foundation of China, the CAS key Project KJCX2-YW-T03, and 973 program 2009CB824800. This work was 
also partially supported by NASA DPR \#NNG08E1671.
J.-M. W. thanks the Natural Science Foundation of China for support via NSFC-10325313, 
10521001 and 10733010. NR is supported by a Ramon y Cajal Fellowship.

\clearpage


\begin{figure*}[t]
\centering
  \includegraphics[angle=270, scale=0.7]{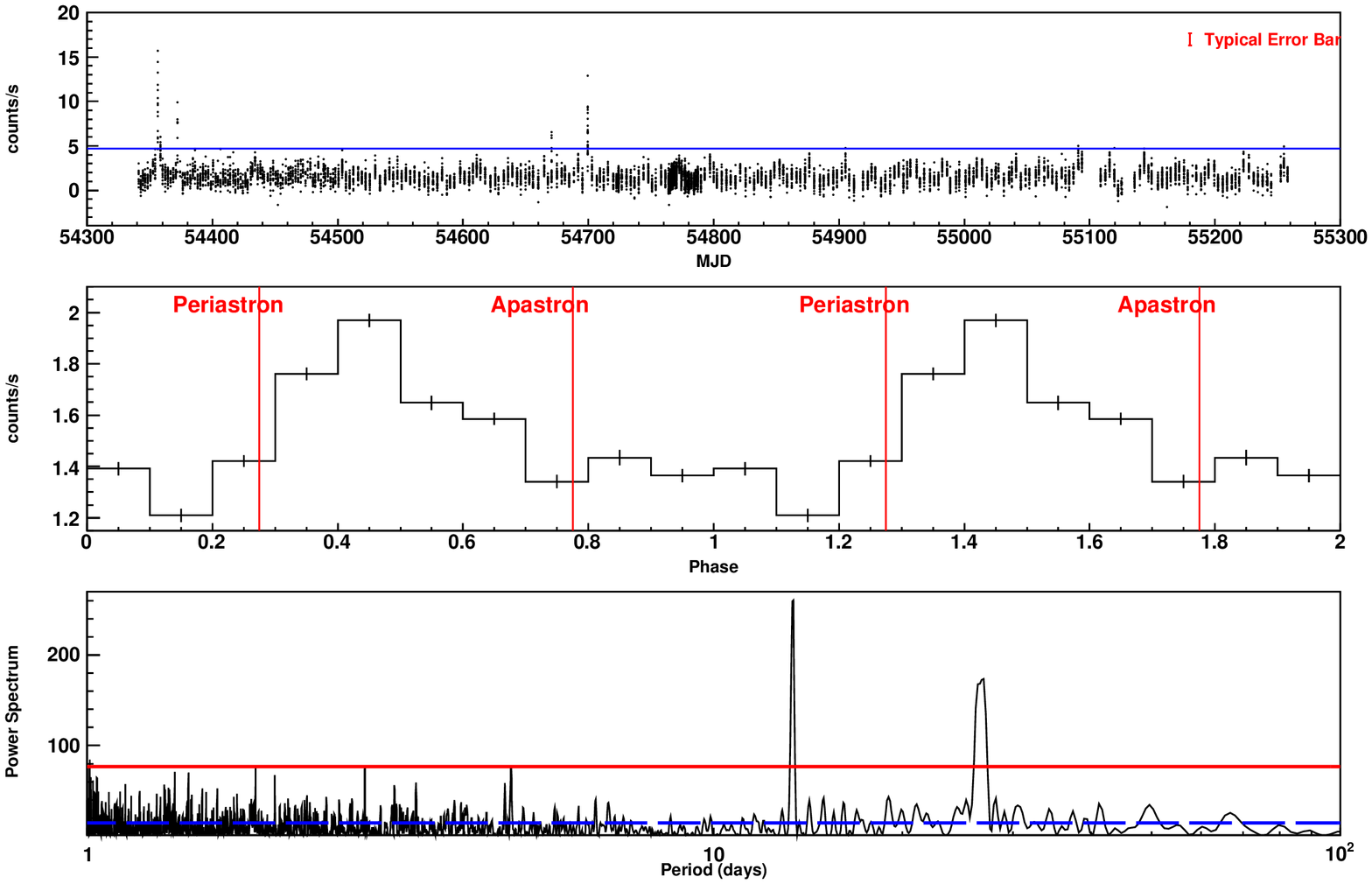}
  \caption{Top: Binned 3--30 keV lightcurve, with 64-second resolution, of the RXTE-PCA data of \lsi\ from 2007 September  to 2010 March. The horizontal line represents the upper flux cut considered in 
our analysis. Middle: Folded lightcurve with orbital ephemeris of \lsi\ using the 
complete RXTE/PCA dataset. Bottom: Power spectrum of the whole RXTE-PCA data. The white 
noise (dashed line) and the red noise (solid line) at the 99\% confidence level are 
plotted in the power spectrum. See text for details.}
  \label{1}
\end{figure*}


\begin{figure*}[t]
\vspace{2cm}
\centering
   \includegraphics[angle=0, scale=0.6]{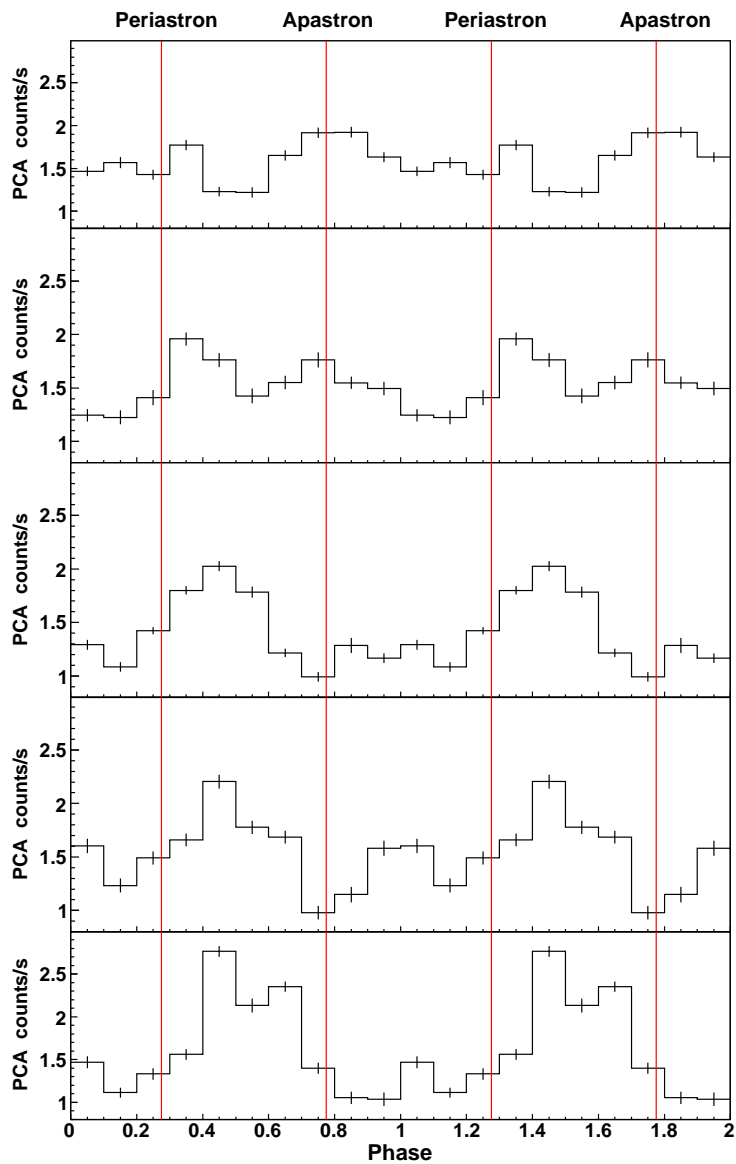}
  \caption{Folded 3--30 keV lightcurves obtained in each of the five separate 6-month 
periods considered. 
  The corresponding dates (dd/mm/yyyy) are: 1st 6-months:  28/8/2007 --- 28/2/2008,   
MJD 54340-54524;
2nd 6-months:  29/2/2008  --- 29/8/2008,   MJD 54525-54707;
3rd 6-months:  30/8/2008  --- 1/3/2009,   MJD 54708-54891;
4th 6-months: 2/3/2009  --- 2/9/2009,   MJD 54892-55076;
5th 6-months:  3/9/2009  --- 3/3/2010,   MJD 55077-55258.}
  \label{6months}
\end{figure*}

\begin{figure*}[t]
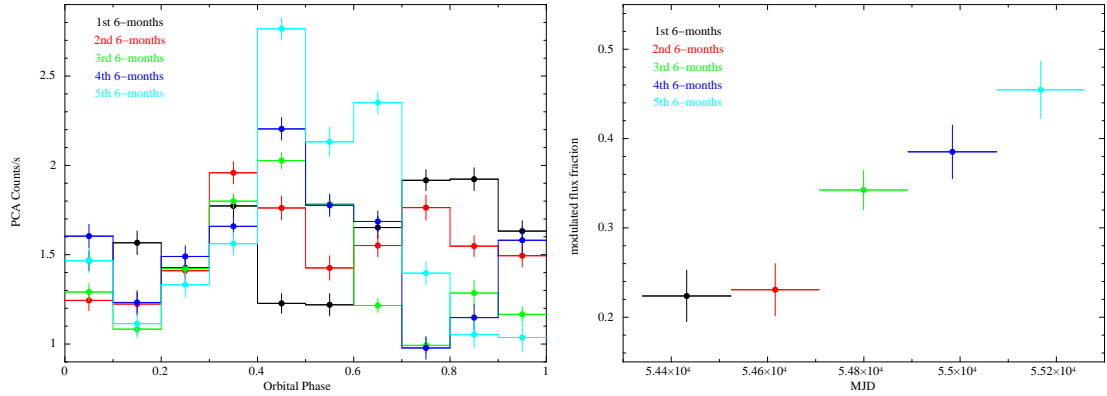

\centering
   \includegraphics[angle=270, scale=0.3]{all_6months_rxte_v2.ps}
     \includegraphics[angle=270, scale=0.3]{orbital_fractions_rxte_v2.ps}
  \caption{Left: 
  Superposition of folded 3--30 keV lightcurves obtained in each of the five separate 6-month 
periods considered. 
  Right: Evolution of the modulated flux fraction in each of the five separate 6-month 
periods considered. 
  The corresponding dates are as in Figure \ref{6months}.}
  \label{2a}
\end{figure*}


\begin{figure*}[t]
\vspace{2cm}
\centering
\rotate
  \includegraphics[angle=0, scale=0.6]{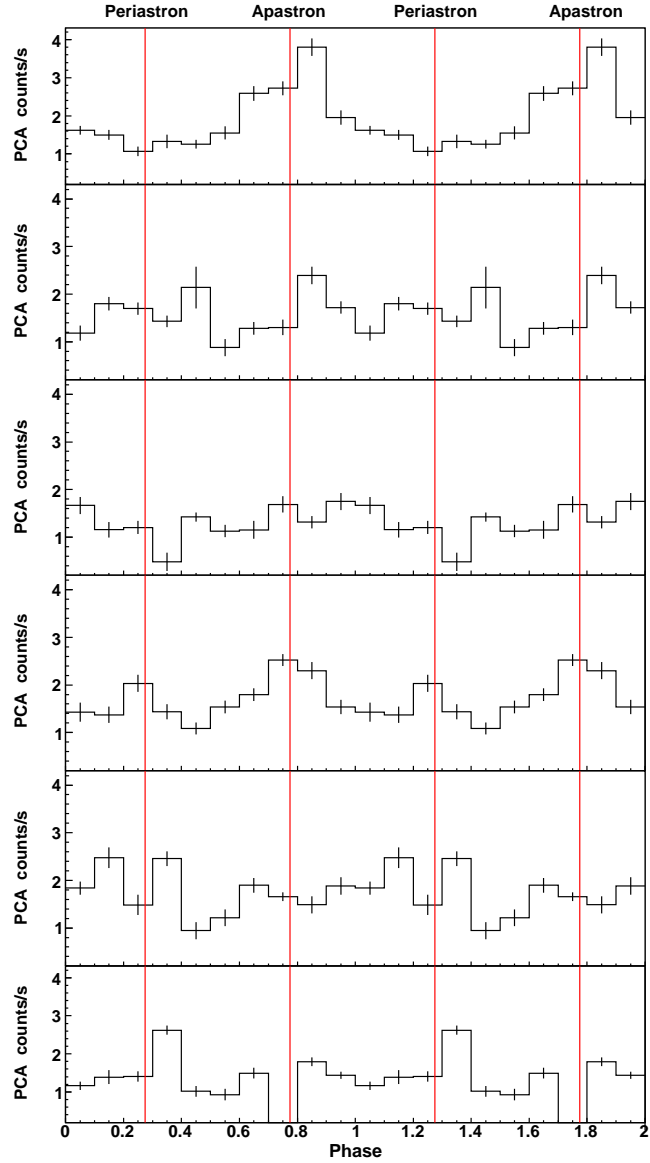}
  \caption{As an example of shorter timescale variability, each panel shows the 3--30 keV lightcurve evolution along each of the months the first 6-months period of data we analyzed, from 28/8/2007 --- 28/2/2008,   
MJD 54340-54524.
Variability in the folded profiles is also visible at these scales. See text for details.}
  \label{22}
\end{figure*}


\end{document}